\documentclass[lettersize,journal]{IEEEtran}
\usepackage{amsthm}
\usepackage{amssymb}
\usepackage{lineno}
\usepackage{graphicx}
\usepackage{latexsym}
\usepackage{amsfonts}
\usepackage{subfigure}
\usepackage{dsfont}
\usepackage{epstopdf}
\usepackage{threeparttable}
\usepackage{multirow}
\usepackage{enumerate}
\usepackage{epsfig}
\usepackage{array}
\usepackage[caption=false,font=normalsize,labelfont=sf,textfont=sf]{subfig}
\usepackage{bm}
 \usepackage{color}
\usepackage{cite}
\usepackage{textcomp}
\usepackage{booktabs}
\usepackage{tabu}
\newcommand{\tabincell}[2]{\begin{tabular}{@{}#1@{}}#2\end{tabular}}
\usepackage{makecell}
\usepackage{multicol}
\usepackage{stfloats}
\usepackage{url}
\usepackage{soul}
\usepackage{verbatim}
\def\BibTeX{{\rm B\kern-.05em{\sc i\kern-.025em b}\kern-.08em
    T\kern-.1667em\lower.7ex\hbox{E}\kern-.125emX}}
\usepackage{amsmath}
\usepackage{algorithm}
\usepackage{algpseudocode}

\usepackage{geometry}
\begin{document}
%
\title{Improved Label Design for Timing Synchronization in OFDM Systems against Multi-path Uncertainty}

\author{Chaojin~Qing,~\IEEEmembership{Member,~IEEE,}
Shuhai~Tang,~Na~Yang,~Chuangui~Rao,~and~Jiafan~Wang
\thanks{This work is supported in part by the Sichuan Science and Technology Program (Grant No. 2021JDRC0003, 23ZDYF0243, 2021YFG0064), the Demonstration Project of Chengdu Major Science and Technology Application (Grant No. 2020-YF09- 00048-SN), the Special Funds of Industry Development of Sichuan Province (Grant No. zyf-2018-056), and the Industry-University Research Innovation Fund of China University (Grant No. 2021ITA10016).}%
\thanks{C. Qing, S. Tang, N. Yang, C. Rao, and J. Wang are with the School of Electrical Engineering and Electronic Information, Xihua University, Chengdu 610039, China (e-mails: qingchj@mail.xhu.edu.cn, tangshh@stu.xhu.edu.cn, yangna6717@163.com, raochuangui5621@163.com, and jifanw@gmail.com).}
}
\maketitle

\begin{abstract}
    Timing synchronization (TS) is vital for orthogonal frequency division multiplexing (OFDM) systems, which makes the discrete Fourier transform (DFT) window start at the inter-symbol-interference (ISI)-free region. However, the multi-path uncertainty in wireless communication scenarios degrades the TS correctness. To alleviate this degradation, we propose a learning-based TS method enhanced by improving the design of training label. In the proposed method, the classic cross-correlator extracts the initial TS feature for benefiting the following machine learning. Wherein, the network architecture unfolds one classic cross-correlation process. Against the multi-path uncertainty, a novel training label is designed by representing the ISI-free region and especially highlighting its approximate midpoint. Therein, the closer to the region boundary of ISI-free the smaller label values are set, expecting to locate the maximum network output in ISI-free region with a high probability. Then, to guarantee the correctness of labeling, we exploit the priori information of line-of-sight (LOS) to form a LOS-aided labeling. Numerical results confirm that, the proposed training label effectively enhances the correctness of the proposed TS learner against the multi-path uncertainty.
\end{abstract}
\IEEEpeerreviewmaketitle
\section{Introduction}
\IEEEPARstart{O}{rthogonal} frequency division multiplexing (OFDM) technology has been widely applied in the modern wireless and mobile communication systems, e.g., the fifth generation (5G) system\cite{ref:5GIoT}. At an OFDM receiver, the correct timing synchronization (TS) locates the starting of discrete Fourier transform (DFT) window per OFDM symbol within its inter-symbol-interference (ISI)-free region \cite{ref:ISIf3}.
However, this task is hardly complete in multi-path propagation scenarios.
In fact, wireless propagation scenarios emerge a lot of multi-path uncertainty, e.g., uncertain multi-path delays and complex path gains, etc  \cite{goldsmith2005wireless}.
Under such uncertainty, the error probability of TS is increased.

To suppress the multi-path uncertainty, the joint TS and channel estimation method via the iterative interference cancellation is proposed in \cite{ref:OMPAlg}.
While this iterative processing requires high computational complexity and large processing delay.
Machine learning, due to its powerful ability in tackling nonlinear problems\cite{MLroad}, can be an alternative way to improve the TS correctness against the impact of multi-path interference.
The authors in \cite{ref:DLITS} investigate a {neural network (NN)}-based signal detection with impacts of TS error and multi-path uncertainty, yet this study does not focus on the TS task.
In a recent study \cite{ref:NewTS}, a convolutional {NN (CNN)}-based TS is investigated, which features a specially designed network architecture to improve the TS correctness. However, this method takes a long time for training and is not conducive to practical application.
In\cite{ref:ELMFTS}, the residual timing offset estimation is accomplished by assuming the achievement of coarse TS and channel estimation, and therefore it omits the impacts of multi-path uncertainty.
Since training labels to be learned are usually helpful to improve the model training without fine-tuning\cite{ref:labELM}, the authors in\cite{ref:LabelTS} improve the TS correctness against multi-path interference by specially designing training labels.
While the uncertainty of multi-path delay is hardly considered in\cite{ref:LabelTS}, which greatly affects the correctness of label designing. Thus, the incorrect labeling limits the improvement of TS correctness provided by \cite{ref:LabelTS}. To our best knowledge, there is limited literature addressing this issue by designing the training label against uncertain multi-path delay.

In this paper, a learning-based TS method aided by the improved label designing is proposed, which aims to improve its adaptability against the multi-path uncertainty.
Specifically, we design novel training label by assigning nonzero values to the label values indexed in the ISI-free region while setting other label values to zeros.
In the designed training label, the values of these labels are set smaller when they are closer to the ISI-free region boundary, which highlights the ISI-free middle region to reduce the risk of timing error.
Nevertheless, the uncertain multi-path delay may result in a variation of ISI-free region with environmental changes, leading to the incorrect labeling and affecting the TS correctness.
To overcome this issue, we further relax the labeling restrictions to line-of-sight (LOS) cases to reserve an enough region for accommodating the uncertain multi-path delay for non-line-of-sight (NLOS) cases.
This avoids the highlighted midpoint being outside the ISI-free region and thus forms the LOS-based priori information.
Without excessively increasing the computational complexity, we combine a single hidden back-propagation {NN} (BPNN) with the classic TS processing to form the learning-based TS (so called TS learner), in which the employed BPNN only unfolds one cross-correlation process.
Numerical results show that the proposed training label can effectively enhance the adaptability and correctness of TS learner against the multi-path uncertainty.

\emph{Notations:} $[\cdot]^T$, $\mathrm{E}\{\cdot\}$, $|\cdot|$, $\lceil\cdot\rceil$, and $({\cdot})^\ast$ denote the operations of transpose, statistical expectation, absolute, ceiling, and complex conjugate, respectively.
\section{System Model}
\begin{figure}[t]
  \centering
  \includegraphics[width=0.49\textwidth]{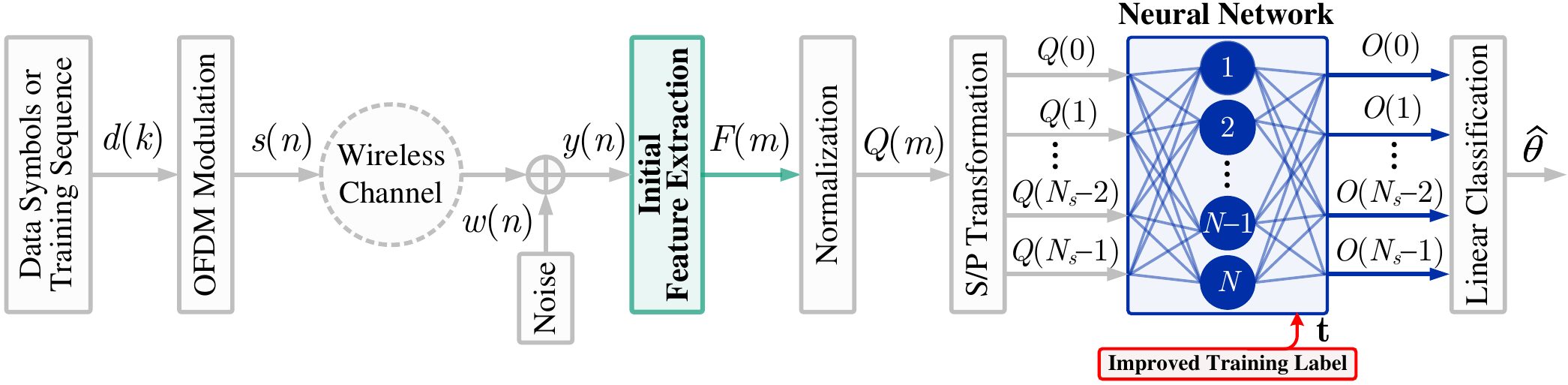}\\
  \caption{System model.}\label{figSS}
\end{figure}
We consider an OFDM system with $N$ subcarriers, as shown in Fig.~\ref{figSS}. At the transmitter, a transmitted OFDM symbol is expressed as
\begin{equation}\label{EQ:1}
s\left( n \right) = \left\{ {\begin{array}{*{20}{l}}
{\sum\limits_{k = 0}^{N - 1} {d\left( k \right)e^ {j \frac{{2\pi k}}{{N}}{n}} } ,}&{0 \le n \le {N}}\\ \vspace{-3mm}\\
{s\left( {n + {N}} \right),}&{ -N_g \le n < 0}
\end{array}} \right.{,}
\end{equation}
where $d(k)$ represents the data symbol or the element of training sequence modulated on the $k$th subcarrier. $N_g$ is the length of cyclic prefix (CP).
In \eqref{EQ:1}, $\mathrm{E}\{|s(n)|^2\}=P_t$ with $P_t$ being the transmitted signal power.
After transmitting $s(n)$ over multi-path channel\cite{ref:CH1}, the received sample is given by
\begin{equation}\label{EQ:3}
  y(n) = e^{j\frac{2\pi n}{N}\varepsilon}\sum\limits^{L}_{l=1}{h_l{{s}}(n-\theta-\tau_l)} +w(n),
\end{equation}
where $\theta$ denotes the timing offset to be estimated, $\varepsilon$ is the normalized carrier frequency offset, and $w(n)$ stands for the additive white Gaussian noise with zero-mean and variance $\sigma^2$.
In \eqref{EQ:3}, $h_l$ and $\tau_l$ are the complex gain and normalized multi-path delay of the $l$-th resolvable path, respectively.
Besides, the multi-path delays are shorten than the CP length to prevent ISI, i.e., $N_g>\tau_l$, and $\tau_l=l-1$ is considered in this paper.

By considering a $N_w$-length observed interval, $N_w$-samples of the received samples are buffered as the observed vector $\mathbf{y}\in\mathbb{C}^{N_w\times1}$, i.e.,
\begin{equation}
\mathbf{y}=\left[y\left(0\right),y\left(1\right),\cdots, y\left(n\right), \cdots, y\left(N_w-1\right)\right]^T,
\end{equation}
where $N_w=2N+N_g$ aims to observe a complete training sequence, and correspondingly, the length of searching range for unknown $\theta$ is $N_s= N_w-N=N+N_g$.
In {Section III}, the proposed TS method for estimating timing offset is elaborated, in which the estimation value of $\theta$ is denoted as $\widehat{\theta}$.
\section{Learning-based TS aided by Label Designing}\label{S:III}

\subsection{Improvement of Label Designing}
By paying special attention to the uncertain multi-path delay, a novel training label $\mathbf{t}\in\mathbb{R}^{N_s\times1}$ is designed to enhance the TS learner, i.e.,
\begin{equation}\label{EQ:5}
{\bf{t}} = {\left[ {\underbrace {0, \cdots ,0}_{\theta  + {\tau _L}},\underbrace {\zeta \left( 1 \right), \cdots , \zeta \left( D \right)}_{\textrm{ISI-free region}},\underbrace {0, \cdots ,0}_{{N_s} - \theta  - {N_g} - 1}} \right]^T},
\end{equation}
where $D=N_g-\tau_L+1$, denoting the length of ISI-free region.
For convenience, a discrete interval for searching $\theta$ is denoted as $\Omega$ and defined as that $ \Omega : \left\{ {m\left| {0 \le m \le {N_s} - 1}, \forall m\in\mathbb{Z} \right.} \right\}$.
Therein, the ISI-free region (denoted as $\Omega_{\textrm{free}}$) is defined as
that ${\Omega _{\textrm{free}}} :  \left\{ {m\left| { {\tau _L} \le m -\theta\le  {N_g}} \right.} \right\}$,
and the ISI region, denoted as $\Omega_{\textrm{ISI}}$, corresponds to a complementary set of ${\Omega _{\textrm{free}}}$ in $\Omega$, i.e., $\Omega_{\textrm{ISI}}\cap \Omega_{\textrm{free}}=\emptyset$ and $\Omega_{\textrm{ISI}}\cup \Omega_{\textrm{free}}=\Omega$.
For $\forall m\in\Omega_{\textrm{free}}$, the $m$-th entry in $\mathbf{t}$ is denoted as $\zeta\left(d\right)$, $d=1,2,\cdots,D$, i.e.,
\begin{equation}\label{EQ:6}
\zeta \left( d \right) = \left\{ {\begin{array}{*{20}{l}}
{d,}&{1\leq d  <  \left\lceil{D+1}/{2}\right\rceil}\\ \vspace{-3mm}\\
{D - d + 1,}&{ \left\lceil{D+1}/{2}\right\rceil  \leq d\leq D}
\end{array}} \right.,
\end{equation}
where each value of $\zeta(d)$ satisfies that the closer to the leftmost or rightmost boundary of $\Omega_{\textrm{free}}$, the smaller values are set.
Different from\cite{ref:LabelTS}, the designed $\mathbf{t}$ in \eqref{EQ:5} not only represents $\Omega_{\textrm{free}}$, but also highlights its middle region.
Since the minimal training loss can be achieved after BPNN training \cite{ref:hecht1992theory}, the maximum value of network output will be concentrated nearby the midpoint of $\Omega_{\textrm{free}}$ with a high probability. That is, the probability of correct TS is increased.

In \eqref{EQ:5}, an approximate midpoint of ${\Omega_{\textrm{free}}}$, denoted as $\mu = \theta+\left\lceil\left(\tau_L+N_g\right)/2\right\rceil$, is considered as the ideal case of correct labeling for \eqref{EQ:5}, i.e., $\mu\in\Omega_\textrm{free}$.
However, due to the uncertain multi-path delay in NLOS cases, ${\mu}\notin\Omega_\textrm{free}$ may appear, which makes error labeling.
If ${\mu}\notin\Omega_\textrm{free}$ has appeared in \eqref{EQ:5}, the trained TS learner will learn the incorrect label, and thus degrades its TS correctness significantly.
To tackle this issue, the labeling restriction is relaxed to LOS cases to reserve the enough region for accommodating the uncertain multi-path delays of NLOS cases.
\subsection{Labeling by LOS-based Priori Information}
By separately denoting $\xi$, $G_r$, $G_t$, and $\lambda$ as the propagation distance, received antenna gain, transmitted antenna gain, and wavelength, the received signal power in a LOS scenario, denoted as $P_r(\xi)$, is given by $ P_r(\xi)=\frac{\lambda^2}{(4\pi\xi)^2}P_tG_tG_r$ \cite{ref:LOSloss}.
With the given transmitted power $P_t$ and the resolvable received power (i.e., a constant value denoted by $P_{\textrm{res}}$), the maximum propagation distance in a LOS scenario (denoted as $\xi_{\textrm{LOS}}$) is ${\xi _{\textrm{LOS}}} = \mathop {\max }\nolimits_\xi  \left\{ {{P_r}\left( \xi  \right) \ge {P_{\textrm{res}}}} \right\}$.
{Therein, the resolvable received power is defined as the received power of the minimal resolvable path during the synchronization phase \cite{ref:thh}.}
For $\forall\xi$ with a given $P_t$, relative to LOS scenarios,
${P_r}\left( \xi  \right)$ of a NLOS scenario is inevitably reduced due to
obstacles\cite{ref:LOSloss}.
Accordingly, with a given $P_t$, the maximum propagation distance in a NLOS scenario, denoted as $\xi_{\textrm{NLOS}}$, satisfies $\xi_{\textrm{NLOS}} \le \xi_{\textrm{LOS}}$.
Correspondingly,
$\tau_{L} \leq \frac{\xi_{\textrm{LOS}}}{c\cdot T}$ is satisfied, where $T$ and $c$ denote the sampling interval and light speed, respectively.
Usually, $\tau_{L}$ of NLOS scenario is difficult to obtain due to the uncertain multi-path delay, while ${\xi_{\textrm{LOS}}}$ seems to be much easier to obtain, so that $\frac{\xi_{\textrm{LOS}}}{c\cdot T}$ can be easily captured.
Without loss of generality, $\frac{\xi_{\textrm{LOS}}}{c\cdot T} < N_g$ is assumed.
{From \cite{hassan2011mathematical}, the CP length is much greater than the maximum propagation delay.
Thus, we explore the prior information of $\frac{\xi_{\textrm{LOS}}}{c\cdot T} < N_g$ to improve the TS correctness.
In fact, this prior information is a loose upper bound, which is only utilized to reflect the heuristic idea of this letter. With the development of integrated sensing and communication (ISAC), the sensing technology (e.g., \cite{zhang2022integration}) can be employed to obtain a tight bound to further improve the TS correctness.}
Then, in a real propagation environment, $\tau_{L} \leq \frac{\xi_{\textrm{LOS}}}{c\cdot T} < N_g$ can be satisfied, yielding
\begin{equation}\label{EQ:QQQWWW}
  \theta+\tau_{L} \leq \theta+\frac{\xi_{\textrm{LOS}}}{c\cdot T} < \theta+ N_g.
\end{equation}
The above \eqref{EQ:QQQWWW} can be viewed as the LOS-based priori information, obtaining a narrowed interval of ISI-free region, i.e.,
\begin{equation}\label{EQ:22222QQQWWW}
\Omega_\textrm{nfree}:\left \{ m| \frac{\xi_{\textrm{LOS}}}{c\cdot T} \leq m-\theta <N_g\right \} s.t.\ \Omega_\textrm{nfree}\subseteq\Omega_\textrm{free}.
\end{equation}
Although the uncertain multi-path delay makes $\mu\in \Omega_\textrm{free}$ hardly, it can be easily achieved by $\mu\in \Omega_\textrm{nfree}$.
Then, $m$ in \eqref{EQ:22222QQQWWW} is substituted by $\mu$, in which $\frac{\xi_{\textrm{LOS}}}{c\cdot T}\le\mu-\theta < N_g$. By using \eqref{EQ:QQQWWW}, we have
\begin{equation}\label{EQ:MidFree}
\theta+\tau_L   \le \mu < \theta+ N_g,
\end{equation}
and thus $\mu\in\Omega_{\textrm{free}}$ is satisfied. By considering $\mu = \theta+\left\lceil\left(\tau_L+N_g\right)/2\right\rceil$, the $\tau_L$ for training can be obtained via $\frac{\xi_{\textrm{LOS}}}{c\cdot T}$.
In \eqref{EQ:QQQWWW}, if $\lceil N_g/2\rceil\le\frac{\xi_{\textrm{LOS}}}{c\cdot T}<N_g$, the $\tau_L$ for training can be set as $2\frac{\xi_{\textrm{LOS}}}{c\cdot T}-N_g\le\tau_L\le\frac{\xi_{\textrm{LOS}}}{c\cdot T}$.
Otherwise, $1\le\tau_L\le\frac{\xi_{\textrm{LOS}}}{c\cdot T}$ can be considered.
For the both cases, a training option of $\tau_{L}$ aided by the LOS priori information of \eqref{EQ:QQQWWW} is expressed as
\begin{equation}\label{EQ:Tautr}
 \left\{ {\begin{array}{*{20}{l}}
{\tau_{L} \sim U\left[1,\left\lceil\frac{\xi_{\textrm{LOS}}}{c\cdot T}\right\rceil\right],}&{\frac{\xi_{\textrm{LOS}}}{c\cdot T}\in\left(0, \left\lceil\frac{N_g}{2}\right\rceil\right)}\\ \vspace{-3mm}\\
{\tau_{L} \sim U\left[2\left\lceil\frac{\xi_{\textrm{LOS}}}{c\cdot T}\right\rceil-N_g,\left\lceil\frac{\xi_{\textrm{LOS}}}{c\cdot T}\right\rceil\right],}&{\frac{\xi_{\textrm{LOS}}}{c\cdot T}\in\left[\left\lceil\frac{N_g}{2}\right\rceil, N_g\right)}\\
\end{array}} \right..
\end{equation}
According to \eqref{EQ:QQQWWW}--\eqref{EQ:Tautr}, the $\tau_{L}$ for training is derived to promise the correct labeling given in \eqref{EQ:5}.
{Since the ISI-free's approximate midpoint $\mu$ is highlighted with $\mu = \theta+\left\lceil\left(\tau_L+N_g\right)/2\right\rceil$ and $\tau_L=2\lceil\mu-\theta\rceil-N_g$, the setting of $\tau$ in \eqref{EQ:Tautr} can guarantee the condition that $\mu \in \Omega_\textrm{nfree}\subseteq\Omega_\textrm{free}$. On the basis of this, the NN trained by the label designed in \eqref{EQ:5} can achieve $\widehat{\theta}\in\Omega_\textrm{free}$ with a high probability. Furthermore, the more cases of $\tau_L$ that are learned, the greater robustness against the varied $\tau_L$ that can be achieved. Therefore, it is reasonable to relax $\tau_L$ to a random value instead of a constant value, as done in \eqref{EQ:Tautr}.}

\vspace{-4mm}
\subsection{Learning-based TS Method}
\begin{table}[t]
\caption{Network Architecture}
\renewcommand\arraystretch{1.25}
\centering
\setlength{\tabcolsep}{2mm}{
\begin{tabu}{c|c|c|c}
\tabucline[0.9pt]{-}
Layer Name & Neuron Nodes &  Activation Function & Normalization \\ \tabucline[0.9pt]{-}
Input Layer      & \(N_s\) &  None    & $\ell_2$-norm         \\ \hline
Hidden Layer& \({N}\)       & Sigmoid        & None \\\hline
Output Layer & \(N_s\)        & None        & None \\
\tabucline[0.9pt]{-}
\end{tabu}}
\label{Tab:S}
  \vspace{-1mm}
\end{table}
\subsubsection{Network Architecture}
The architecture of the designed TS learner is illustrated in TABLE~\ref{Tab:S}, which has an input layer, a hidden layer and an output layer.
To avoid excessively increasing complexity, the neurons of input layer, hidden layer, and output layer are set as $N_s$, $N$, and $N_s$, respectively. Therein, the $N$-neuron hidden layer is derived from considering unfolding one cross-correlation process.
{Since the input scale of NN may differ from those of other layers, it is practical to employ the $\ell_2$-norm in the input layer.
Besides, the hidden layer employs the sigmoid activation function, i.e., $f(x)=1/(1+e^{-x})$,
for the reason that it is easy to calculate and suitable for shallow NNs \cite{8407425}.}
\subsubsection{Initial Feature Extraction}
By employing the classic TS method, its timing metric is utilized as the initial $N_s$ features of TS to facilitate the model learning\cite{RepLearn}, i.e.,
\begin{equation}\label{EQ:11}
F\left( m \right) = \left| {\sum\limits_{n = 0}^{N - 1} {{{x}^\ast}\left( n \right)y\left( {m + n} \right)} } \right|^2, 0\le m\le N_s-1,
\end{equation}
{where $x(n)$ is a local training sequence, e.g., the Zadoff-Chu sequence in \cite{ref:ZCseq}.}
{By using the $\ell_2$-norm of $F(m)$, the network input, denoted as $Q(m)$, is normalized to the interval $\left[0,1\right)$, i.e., $0\le Q(m)<1$, which is expressed as $Q(m)={F(m)}/{\sqrt{\sum\nolimits^{N_s-1}_{m=0}{|F(m)|^2}}}$.}
The vector form of $Q(m)$, denoted as $\mathbf{Q}\in\mathbb{R}^{N_s\times1}$, is expressed as ${\bf{Q}} = {\left[ {Q\left( 0 \right), \cdots ,Q\left( {{N_s} - 1} \right)} \right]^T}$.
\subsubsection{Offline Training}
First, by using \eqref{EQ:1}--\eqref{EQ:11}, the training data set $\{\mathbf{Q}_i,\mathbf{t}_i\}^{N_t=10^4}_{i=1}$ are obtained.
For the $i$-th training sample, we employ the exponentially decayed channel model with its decayed factor $\eta_i$ being $\eta_i\sim U[0.01,0.5]$ for combating the uncertainty of complex path gain.
Meanwhile, $\frac{\xi_{\textrm{LOS}}}{c\cdot T}=\lceil7N_g/8\rceil$ is given, which derives $\tau_{L,i}\sim U[\lceil3N_g/4\rceil,\lceil7N_g/8\rceil]$ for the label designing in \eqref{EQ:5} to combat the uncertain multi-path delay.
{Besides, the SNR for generating the $i$-th training sample is randomly selected from {$\{-2\mathrm{dB}, 0\mathrm{dB}, 2\mathrm{dB}, ... , 10\mathrm{dB}\}$} with a probability of 1/7}.

In the developed TS learner, the optimizer employs the stochastic gradient descent (SGD) algorithm and sets the initial learning rate as $\alpha=0.001$ \cite{ref:adam}.
{In this paper, the loss function is defined as
\begin{equation}\label{Loss}
  \emph{L}_{\bf \Theta} = \frac{1}{N_t}\sum\limits_{i =1}^{N_t} {\left\| {{G_{{\bf \Theta}}}\left( {{{\bf{Q }}_i}} \right) - {{\bf{t}}_i}} \right\|_2^2},
\end{equation}
where $\bf \Theta$ is a set of network parameters (i.e., weights and biases) to be optimized, and $G_{{\bf \Theta}}(\cdot)$ is a mapping function parameterized by ${\bf \Theta}$.
By setting $J$ as the total iterative steeps, the network optimization is defined as\cite{sgd}
\begin{equation}\label{EQ:SGD}
{{{\bf \Theta} _{j + 1}} \leftarrow {{\bf \Theta}_j} - \alpha \nabla \emph{L}_{{\bf \Theta}_j}},
\end{equation}
where ${\bf \Theta}_j$, $j=1,2,\cdots,J$, denotes the network parameters after the $j$th optimization, and $\nabla\emph{L}_{{\bf \Theta}_j}$ is the gradient of $\emph{L}_{{\bf \Theta}_j}$.}

\subsubsection{Online Deployment}
According to \eqref{EQ:1}--\eqref{EQ:3} and \eqref{EQ:11}, the initial TS features are first extracted, and then normalized to form the network input $\mathbf{Q}$.
With the trained $G_{{\bf \Theta}}$,
the network output $\mathbf{O}\in\mathbb{R}^{N_s\times1}$ is obtained by ${\bf O} = G_{{\bf \Theta}}\left({\bf Q}\right)$.
Finally, the output $\mathbf{O}$ is expressed as $\left[O(0),\cdots,O(m),\cdots,O(N_s-1)\right]^T$ for timing offset estimation, i.e., $\widehat{\theta}  = \mathop {\arg\max }\limits_{ m\in \Omega} |{O\left( m \right)}|$.
\section{Numerical Results}

\subsection{Parameter Setting}
\begin{figure}[t]
  \vspace{-3mm}
  \centering
  \includegraphics[width=0.40\textwidth]{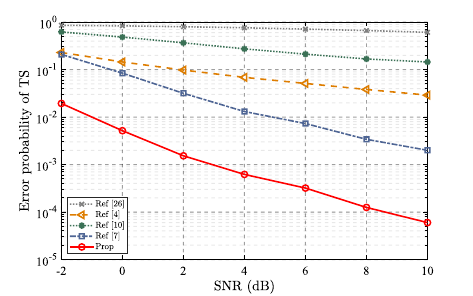}\\
  \caption{Effectiveness analysis.}\label{figeff}
\end{figure}

In the simulations, we consider that $N=128$, $N_g=32$, $N_w=288$, and $N_s=160$.
The error probability of TS is utilized to evaluate the TS correctness, which is defined as the probability of estimating the timing offset outside of the ISI-free region.
{The OFDM technology is primarily employed to combat the ISI caused by the multi-path propagation \cite{Ref:OFDMsys}, and thereby the frequency selective fading channels are mainly considered in the simulations.}
{We leverage the delay spread profile to quantify the uncertainty of multi-path fading \cite{telatar2000capacity}. To simulate the multi-path uncertainty, Fig.~\ref{figeff} to Fig.~\ref{FiggenCH} depict the TS correctness of the proposed method for the cases where the training delay spread profile differs from the testing ones.}
{The case of the maximum LOS propagation delay (i.e., $\frac{\xi_{\textrm{LOS}}}{c\cdot T}$) is set only to provide the LOS priori information (i.e., \eqref{EQ:QQQWWW}) for assisting the improvement of label designing in \eqref{EQ:5}.
Consequently, we consider a relatively large value of $\frac{\xi_{\textrm{LOS}}}{c\cdot T}$ to guarantee the correctness of LOS priori information, i.e., $\frac{\xi_{\textrm{LOS}}}{c\cdot T}$ is set as $\lceil7N_g/8\rceil$.}
In the simulations, the exponentially decayed Rayleigh fading channel in \cite{ref:CH1} and different 5G tapped-delay-line (TDL) channel models (e.g., TDL-B and TDL-C given in 3GPP TR 38.901 \cite{ref:5Gtdl}) are employed.
{Besides, the signal-to-noise ratio (SNR) is defined as ${\rm SNR} = 10\log_{10}(P_t/\sigma^2_n)$\cite{SNRdefine}, and correspondingly, $\sum^L_{l=1}|h_l|^2=1$ is considered \cite{ref:CH1}.}

For the ease of description, the TS learner with the training labels proposed in this paper and \cite{ref:LabelTS} are referred as to ``Prop'' and ``Ref \cite{ref:LabelTS}'', respectively. ``Ref \cite{ref:NewTS}'' is the TS method given in \cite{ref:NewTS}. ``Ref \cite{ref:OMPAlg}'' stands for the iterative-based TS method in \cite{ref:OMPAlg}.
Besides, the classic TS method in \cite{ref:CTS}, denoted as ``Ref\cite{ref:CTS}'', serves as the baseline.

\subsection{Computational Complexity and Processing Delay}
\begin{table}[t]
\renewcommand{\arraystretch}{1.25}
\caption{Computational Complexity and Processing Delay}
\label{table_I}
\centering
\setlength{\tabcolsep}{1.25mm}{
\begin{tabu}{c|c|c|c}
\tabucline[1 pt]{-}
 Method      & Computational Complexity  & \tabincell{c}{Example\vspace{-0.5mm}\\(CM)} & \tabincell{c}{Processing \vspace{-0.5mm}\\Delay (sec)} \\
 \tabucline[1 pt]{-}
    Ref \cite{ref:OMPAlg}   & 
    {$LN{N_s} + \sum\nolimits_{l = 1}^L ({3l{N_s}}  + {l^3} + {l^2}{N_s})$}              &  2167396   & 81.41         \\ \hline
    Ref \cite{ref:LabelTS}   & $1.5N+4(N_s-1)+16N_s^2$              &  410428 & 1.614\\  \hline
    Ref \cite{ref:NewTS}     & ${0.5N_s^2+2NN_s+1.5N_gN_s+N_s}$              &  {{70240}}  &{0.349}        \\\hline
    Prop     & ${NN_s+0.5NN_s}$              &  {{30720}}  &{0.141}        \\
    \tabucline[1 pt]{-}
\end{tabu}
}
\end{table}
\begin{figure}[t]
  \centering
  \includegraphics[width=0.42\textwidth]{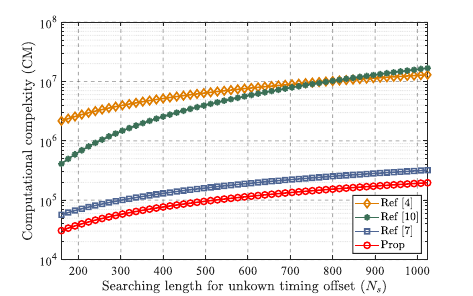}\\
  \caption{Computational complexity analysis.}\label{figCM}
\end{figure}
{TABLE~\ref{table_I} presents the expression of computational complexity, in which the complex multiplication (CM) is used for evaluating the computational complexity.
By considering the impact of searching length $N_s$, Fig.~\ref{figCM} plots the CM of each given TS method in terms of $N_s$, where $N_s$ increases from 160 to 1024 with the interval being 16.
All the evaluations are carried out on an Intel Core i5-11300H, 3.10GHz CPU, and $L=28$ is considered for $10^4$ experiments in both TABLE~\ref{table_I} and Fig.~\ref{figCM}.
In Fig.~\ref{figCM}, the CM of ``Prop'' is smaller than those of ``Ref\cite{ref:OMPAlg}'', ``Ref \cite{ref:NewTS}'', and ``Ref\cite{ref:LabelTS}''.
Similarly, TABLE~\ref{table_I} reflects that ``Prop'' obtains a lower computational complexity and processing delay, and the case where $N_s=160$ is given.
The reason is that ``Prop'' unfolds one cross-correlation process, while others unfold at least two iterations of the cross-correlation process.
From TABLE~\ref{Tab:S}, the CM of the designed NN of ``Prop'' is $0.5NN_s$, which does not exceed the CM of one cross-correlation process, i.e., $NN_s$. Naturally, the CM of ``Prop'' will not exceed the CM of two cross-correlation process.
Thus, the relatively lightweight NN is employed by ``Prop'' compared with the given TS methods.}

\subsection{Effectiveness Analysis}
Fig.~\ref{figeff} plots the error probability of TS to evaluate the TS performance of different methods against multi-path interference and its uncertain multi-path delay.
Meanwhile, the case of $\{\tau_L=27,\eta=\ln10/(L-1)\}$is employed to simulate the uncertainty of multi-path delay, while the case of $\{\tau_L\sim U[24,28],\eta\sim U[0.01,0.5]\}$ is utilized for the offline training presented in {Section III-C}.
For each given SNR in Fig.~\ref{figeff}, ``Prop'' obtains significantly smaller error probability of TS compared with other given TS methods.
{Compared with ``Prop'', ``Ref \cite{ref:CTS}'' presents a weaker robustness against the increased numbers of multi-path due to the lack of the consideration of rich multi-path scenarios (only 3 or 6 taps are considered in the simulations in \cite{ref:CTS}).}
Relative to ``Ref\cite{ref:LabelTS}'', the superiority of ``Prop'' is that the proposed training label is not only specially designed for decreasing the risk of estimating the timing offset outside the ISI-free region, but also improved by LOS-based labeling for increasing the correctness of labeling.
Although ``Prop'' and ``Ref \cite{ref:NewTS}'' both exploit the LOS-based priori information for correct labeling, the training label designed by ``Prop'' is superior due to the highlighting of ISI-free region and its approximate middle point.
{This reveals the feasibility of enhancing the label designing in improving the TS correctness.}
Last but not least, compared with ``Ref \cite{ref:OMPAlg}'', ``Prop'' achieves a lower error probability with the reduced computational complexity, due to the powerful ability of machine learning in coping with the deficiencies such as noise and multi-path interference.
{This reflects that the learning-based TS method with an appropriate training label can improve the TS correctness with the reduced complexity, and thereby the proposed method provides an alternative solution for alleviating the demand of complexly modeling in a practical system.}
To sum up, relative to ``Ref\cite{ref:CTS}'', ``Ref\cite{ref:OMPAlg}'', ``Ref \cite{ref:NewTS}'', and ``Ref\cite{ref:LabelTS}'', ``Prop'' can effectively improves the TS correctness against uncertain multi-path delay.
\begin{figure}[t]
  \centering
  \includegraphics[width=0.40\textwidth]{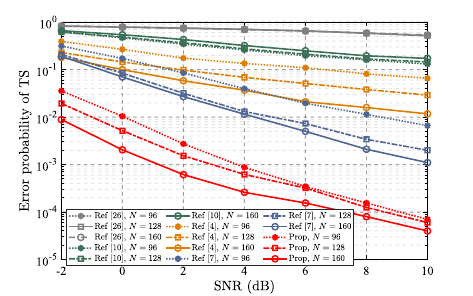}\\
  \caption{Robustness analysis.}\label{figeffN}
\end{figure}
\begin{figure}[t]
\centering
\includegraphics[width=0.40\textwidth]{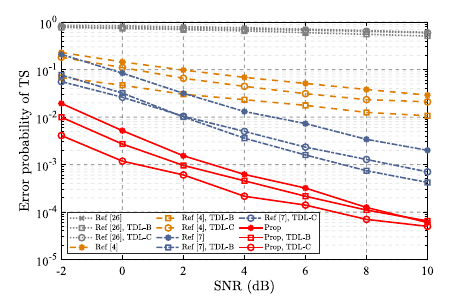}\\
\caption{Generalization analysis.}\label{FiggenCH}
\end{figure}
\subsection{Robustness Analysis}
{To analyze the robustness against the training sequence length (i.e., $N$), Fig.~\ref{figeffN} plots the error probability of TS, where $N=96$, $N=128$, and $N=160$ are considered. Except for $N$, other parameters remain the same as those mentioned in Section III-C.
For each given value of $N$ in Fig.~\ref{figeffN}, ``Prop'' reaches the smallest error probability compared with other given TS methods. This reflects the robustness of ``Prop'' against the change of $N$. Meanwhile, it is worth noting that the error probability of ``Prop'' reduces with the increases of $N$. This is due to the fact that the ability of anti-noise in cross-correlator is enhanced (deteriorated) with the increase (decrease) of $N$.
Therefore, the proposed learning-based TS method assisted by the improved label design can robustly improve the TS correctness.}

\subsection{Generalization Analysis}
Fig.~\ref{FiggenCH} presents the comparison of error probability in different channel models to analyze the generalization performance of ``Prop''. Except for the testing wireless channel models, other parameters are the same as those mentioned in Section IV-B. {Importantly, the NN adopted in this letter dose not need to be retrained when
the testing channel model differs from the training one}.
For each given channel model in Fig.~\ref{FiggenCH}, ``Prop'' reaches the lowest error probability among the given TS methods.
Furthermore, it is not obvious that, for all given SNRs, the fluctuations of the error probabilities of ``Prop'' caused by different channel models.
Therefore, the TS correctness of ``Prop'' is superior to those of other TS methods.
This shows that the proposed TS methods possesses a good generalization performance against different 5G TDL channel models.
\section{Conclusion}
Against the influence of uncertain multi-path delay, the proposed learning-based TS method aided by the improved label designing is investigated in OFDM systems.
By highlighting the ISI-free region and its the approximate midpoint against uncertain multi-path delay, the designed training label effectively reduces the risk that the DFT window starts at the ISI region.
Meanwhile, with the LOS-based priori information, the incorrect labeling affected by the uncertain multi-path delay is further rectified.
Simulation results validates the effectiveness and generalization of designed training label in improving the TS correctness of the learning-based TS method against multi-path uncertainty.
{In our future works, we will investigate the the sensing-aided TS from the perspective of ISAC.}
\vspace{-5mm}
\bibliographystyle{ieeetran}
\bibliography{ref}
\end{document}